
\input epsf
%
%
\catcode`\@=11 
\newcount\yearltd\yearltd=\year\advance\yearltd by -1900
%

\def\draftmode{\message{ DRAFTMODE }\def\draftdate{{\rm preliminary draft:
\number\month/\number\day/\number\yearltd\ \ \hourmin}}%
\headline={\hfil\draftdate}\writelabels\baselineskip=20pt plus 2pt minus 2pt
 {\count255=\time\divide\count255 by 60 \xdef\hourmin{\number\count255}
  \multiply\count255 by-60\advance\count255 by\time
  \xdef\hourmin{\hourmin:\ifnum\count255<10 0\fi\the\count255}}}
\def\nolabels{\def\wrlabeL##1{}\def\eqlabeL##1{}\def\reflabeL##1{}}
\def\writelabels{\def\wrlabeL##1{\leavevmode\vadjust{\rlap{\smash%
{\line{{\escapechar=` \hfill\rlap{\sevenrm\hskip.03in\string##1}}}}}}}%
\def\eqlabeL##1{{\escapechar-1\rlap{\sevenrm\hskip.05in\string##1}}}%
\def\reflabeL##1{\noexpand\llap{\noexpand\sevenrm\string\string\string##1}}}
\nolabels
%
\global\newcount\secno \global\secno=0
\global\newcount\meqno \global\meqno=1
\def\newsec#1{\global\advance\secno by1\message{(\the\secno. #1)}
\global\subsecno=0\eqnres@t\noindent{\bf\the\secno ~#1}
\writetoca{{\secsym} {#1}}\par\nobreak\medskip\nobreak}
\def\eqnres@t{\xdef\secsym{\the\secno.}\global\meqno=1\bigbreak\bigskip}
\def\sequentialequations{\def\eqnres@t{\bigbreak}}\xdef\secsym{}
\global\newcount\subsecno \global\subsecno=0
\def\subsec#1{\global\advance\subsecno by1\message{(\secsym\the\subsecno. #1)}
\ifnum\lastpenalty>9000\else\bigbreak\fi
\noindent{\it\secsym\the\subsecno ~#1}\writetoca{\string\quad
{\secsym\the\subsecno.} {#1}}\par\nobreak\medskip\nobreak}
\def\appendix#1{\global\meqno=1\global\subsecno=0\xdef\secsym{\hbox{#1}}
\bigbreak\bigskip\noindent{\bf #1}
\writetoca{{#1}}\par\nobreak\smallskip\nobreak}
%
%
\def\eqnn#1{\xdef #1{(\secsym\the\meqno)}\writedef{#1\leftbracket#1}%
\global\advance\meqno by1\wrlabeL#1}
\def\eqna#1{\xdef #1##1{\hbox{$(\secsym\the\meqno##1)$}}
\writedef{#1\numbersign1\leftbracket#1{\numbersign1}}%
\global\advance\meqno by1\wrlabeL{#1$\{\}$}}
\def\eqn#1#2{\xdef #1{(\secsym\the\meqno)}\writedef{#1\leftbracket#1}%
\global\advance\meqno by1$$#2\eqno#1\eqlabeL#1$$}
%
%
\global\newcount\refno \global\refno=1
\newwrite\rfile
\def\ref{$^{\the\refno}$\nref}
\def\nref#1{\xdef#1{\the\refno.}\writedef{#1\leftbracket#1}%
\ifnum\refno=1\immediate\openout\rfile=refs.tmp\fi
\global\advance\refno by1\chardef\wfile=\rfile\immediate
\write\rfile{\noexpand\item{#1\ }\reflabeL{#1\hskip.31in}\pctsign}\findarg}
\def\findarg#1#{\begingroup\obeylines\newlinechar=`\^^M\pass@rg}
{\obeylines\gdef\pass@rg#1{\writ@line\relax #1^^M\hbox{}^^M}%
\gdef\writ@line#1^^M{\expandafter\toks0\expandafter{\striprel@x #1}%
\edef\next{\the\toks0}\ifx\next\em@rk\let\next=\endgroup\else\ifx\next\empty%
\else\immediate\write\wfile{\the\toks0}\fi\let\next=\writ@line\fi\next\relax}}
\def\striprel@x#1{} \def\em@rk{\hbox{}}
\def\lref{\begingroup\obeylines\lr@f}
\def\lr@f#1#2{\gdef#1{\ref#1{#2}}\endgroup\unskip}

\def\addref#1{\immediate\write\rfile{\noexpand\item{}#1}} 
\def\startrefs#1{\immediate\openout\rfile=refs.tmp\refno=#1}
\def\xref{\expandafter\xr@f}\def\xr@f#1.{#1}
\def\cite{\expandafter\cxr@f}\def\cxr@f#1.{$^{#1}$}
\def\xcite{\expandafter\xcxr@f}\def\xcxr@f#1.{{#1}}
\def\cites#1{\count255=1$^{\r@fs #1{\hbox{}}}$}
\def\r@fs#1{\ifx\und@fined#1\message{reflabel \string#1 is undefined.}%
\nref#1{need to supply reference \string#1.}\fi%
\vphantom{\hphantom{#1}}\edef\next{#1}\ifx\next\em@rk\def\next{}%
\else\ifx\next#1\ifodd\count255\relax\xref#1\count255=0\fi%
\else#1\count255=1\fi\let\next=\r@fs\fi\next}
\newwrite\lfile
{\escapechar-1\xdef\pctsign{\string\%}\xdef\leftbracket{\string\{}
\xdef\rightbracket{\string\}}\xdef\numbersign{\string\#}}

\def\writestop{\def\writestoppt{\immediate\write\lfile{\string\pageno%
\the\pageno\string\startrefs\leftbracket\the\refno\rightbracket%
\string\def\string\secsym\leftbracket\secsym\rightbracket%
\string\secno\the\secno\string\meqno\the\meqno}\immediate\closeout\lfile}}
\def\writestoppt{}\def\writedef#1{}
\def\seclab#1{\xdef #1{\the\secno}\writedef{#1\leftbracket#1}\wrlabeL{#1=#1}}
\def\subseclab#1{\xdef #1{\secsym\the\subsecno}%
\writedef{#1\leftbracket#1}\wrlabeL{#1=#1}}
\newwrite\tfile \def\writetoca#1{}
\def\leaderfill{\leaders\hbox to 1em{\hss.\hss}\hfill}
\def\writetoc{\immediate\openout\tfile=toc.tmp
   \def\writetoca##1{{\edef\next{\write\tfile{\noindent ##1
   \string\leaderfill {\noexpand\number\pageno} \par}}\next}}}

%
\catcode`\@=12 
%
%
\font\abssl=cmsl10 scaled 833
\font\absrm=cmr10 scaled 833 \font\absrms=cmr7 scaled  833
\font\absrmss=cmr5 scaled  833 \font\absi=cmmi10 scaled  833
\font\absis=cmmi7 scaled  833 \font\absiss=cmmi5 scaled  833
\font\abssy=cmsy10 scaled  833 \font\abssys=cmsy7 scaled  833
\font\abssyss=cmsy5 scaled  833 \font\absbf=cmbx10 scaled 833
\skewchar\absi='177 \skewchar\absis='177 \skewchar\absiss='177
\skewchar\abssy='60 \skewchar\abssys='60 \skewchar\abssyss='60
\def\abstractfont{\def\rm{\fam0\absrm}
\textfont0=\absrm \scriptfont0=\absrms \scriptscriptfont0=\absrmss
\textfont1=\absi \scriptfont1=\absis \scriptscriptfont1=\absiss
\textfont2=\abssy \scriptfont2=\abssys \scriptscriptfont2=\abssyss
\textfont\itfam=\absi \def\it{\fam\itfam\absi}
\textfont\slfam=\abssl \def\sl{\fam\slfam\abssl}
\textfont\bffam=\absbf \def\bf{\fam\bffam\absbf}\rm}
\font\ftsl=cmsl10 scaled 833
\font\ftrm=cmr10 scaled 833 \font\ftrms=cmr7 scaled  833
\font\ftrmss=cmr5 scaled  833 \font\fti=cmmi10 scaled  833
\font\ftis=cmmi7 scaled  833 \font\ftiss=cmmi5 scaled  833
\font\ftsy=cmsy10 scaled  833 \font\ftsys=cmsy7 scaled  833
\font\ftsyss=cmsy5 scaled  833 \font\ftbf=cmbx10 scaled 833
\skewchar\fti='177 \skewchar\ftis='177 \skewchar\ftiss='177
\skewchar\ftsy='60 \skewchar\ftsys='60 \skewchar\ftsyss='60
\def\footnotefont{\def\rm{\fam0\ftrm}
\textfont0=\ftrm \scriptfont0=\ftrms \scriptscriptfont0=\ftrmss
\textfont1=\fti \scriptfont1=\ftis \scriptscriptfont1=\ftiss
\textfont2=\ftsy \scriptfont2=\ftsys \scriptscriptfont2=\ftsyss
\textfont\itfam=\fti \def\it{\fam\itfam\fti}%
\textfont\slfam=\ftsl \def\sl{\fam\slfam\ftsl}%
\textfont\bffam=\ftbf \def\bf{\fam\bffam\ftbf}\rm}
\font\ninerm=cmr9 \font\sixrm=cmr6 \font\ninei=cmmi9 \font\sixi=cmmi6
\font\ninesy=cmsy9 \font\sixsy=cmsy6 \font\ninebf=cmbx9
\font\nineit=cmti9 \font\ninesl=cmsl9 \skewchar\ninei='177
\skewchar\sixi='177 \skewchar\ninesy='60 \skewchar\sixsy='60
\def\ninepoint{\def\rm{\fam0\ninerm}
\textfont0=\ninerm \scriptfont0=\sixrm \scriptscriptfont0=\fiverm
\textfont1=\ninei \scriptfont1=\sixi \scriptscriptfont1=\fivei
\textfont2=\ninesy \scriptfont2=\sixsy \scriptscriptfont2=\fivesy
\textfont\itfam=\ninei \def\it{\fam\itfam\nineit}\def\sl{\fam\slfam\ninesl}%
\textfont\bffam=\ninebf \def\bf{\fam\bffam\ninebf}\rm}
%
%

\vsize=7.0truein
\hsize=4.7truein
\baselineskip 12truept plus 0.5truept minus 0.5truept
\hoffset=0.5truein
\voffset=0.5truein

\def\gsim{\mathrel{\rlap{\lower4pt\hbox{\hskip1pt$\sim$}}
    \raise1pt\hbox{$>$}}}         

\def\frac#1#2{{{#1}\over {#2}}}
\def\half{\hbox{${1\over 2}$}}

\catcode`@=11 
\def\slash#1{\mathord{\mathpalette\c@ncel#1}}
 \def\c@ncel#1#2{\ooalign{$\hfil#1\mkern1mu/\hfil$\crcr$#1#2$}}
\def\lsim{\mathrel{\mathpalette\@versim<}}
\def\gsim{\mathrel{\mathpalette\@versim>}}
 \def\@versim#1#2{\lower0.2ex\vbox{\baselineskip\z@skip\lineskip\z@skip
       \lineskiplimit\z@\ialign{$\m@th#1\hfil##$\crcr#2\crcr\sim\crcr}}}
\catcode`@=12 

\def\NP{{\it Nucl.~Phys.~}}

\def\PL{{\it Phys.~Lett.~}}

\def\ZP{{\it Zeit.~Phys.~}}

\def\vol#1{{\bf #1}}\def\vyp#1#2#3{\vol{#1}, #3 (#2)}


\tolerance=10000
\hfuzz=5pt
\pageno=0\nopagenumbers\tolerance=10000\hfuzz=5pt
\line{\hfill {\tt hep-ph/9805315}}
\line{\hfill Edinburgh 98/6}
\line{\hfill DFTT 22/98}
\vskip 24pt
\centerline{\bf CORRECTIONS AT SMALL X}
\vskip 36pt\centerline{Richard D. Ball\footnote*{\footnotefont
Royal Society University Research Fellow}}
\vskip 12pt
\centerline{\it Department of Physics and Astronomy}
\centerline{\it University of Edinburgh, EH9 3JZ, Scotland}
\vskip 12pt\centerline{and}
\vskip 12pt\centerline{Stefano Forte}
\vskip 12pt
\centerline{\it INFN, Sezione di Torino, Via P. Giuria 1}
\centerline{\it I-10125 Torino, Italy}
\vskip 48pt
{\narrower\baselineskip 10pt
\centerline{\bf Abstract}
\medskip\noindent
We show that the all order summation of small x logarithms 
cannot be included in the $Q^2$  evolution of structure
functions because the NLLx corrections overwhelm 
the LLx contribution.
\smallskip}
\bigskip
\centerline{Talk at the }
\centerline{\it `6th International Workshop on Deep Inelastic Scattering and QCD'}
\centerline{{\it (DIS98)}, Brussels, April 1998}
\medskip
\centerline{\it to be published in the proceedings}
\vskip 55pt
\line{May 1998\hfill}
\eject
\footline={\hss\tenrm\folio\hss}
\centerline{\bf CORRECTIONS AT SMALL X}
\bigskip\bigskip
{\ninepoint
\centerline{R.D.~BALL}
\smallskip
\centerline{\it Department of Physics and Astronomy,}
\centerline{\it University of Edinburgh, EH9 3JZ, Scotland}
\smallskip\centerline{and}
\smallskip\centerline{S.~FORTE}
\centerline{\it INFN, Sezione di Torino, Via P. Giuria 1}
\centerline{\it I-10125 Torino, Italy}
}
\bigskip
{\abstractfont\baselineskip 9 pt
\advance\leftskip by 36truept\advance\rightskip by 36truept\noindent
We show that the all order summation of small x logarithms 
cannot be included in the $Q^2$  evolution of structure
functions because the NLLx corrections overwhelm 
the LLx contribution.
\smallskip}

\baselineskip 12pt plus 0.5pt minus 0.5pt
\bigskip\bigskip
\goodbreak


\def\sqr#1#2{{\vcenter{\vbox{\hrule height.#2pt
\hbox{\vrule width.#2pt height#1pt \kern#1pt \vrule width.#2pt}
\hrule height.#2pt}}}}
\def\square{\mathchoice\sqr34\sqr34\sqr{2.1}3\sqr{1.5}3}

\nref\FL{V.S.~Fadin and L.N.~Lipatov, {\tt hep-ph/9802290} and in
these proceedings.}
\nref\CC{M.~Ciafaloni and G.~Camici,  {\tt hep-ph/9803389} and in
these proceedings.}
\nref\CH{S.~Catani and F.~Hautmann, \NP\vyp{B427}{1994}{475}.}
\nref\Summing{R.D.~Ball and S.~Forte, \PL\vyp{B351}{1995}{313}.} 
\nref\Scheme{S.~Forte and R.D.~Ball, {\tt hep-ph/9507211} (DIS95 proceedings).}
\nref\CQz{M.~Ciafaloni, \PL\vyp{B356}{1995}{74}; \NP\vyp{B496}{1997}{305}.}
\nref\SDIS{S.~Catani, \ZP\vyp{C70}{1996}{263}.}
\nref\GDIS{S.~Catani, {\tt hep-ph/9608310} (DIS96 proceedings).}
\nref\Mom{R.D.~Ball and S.~Forte, \PL\vyp{B359}{1995}{362}.}
\nref\Rome{S.~Forte and R.D.~Ball, {\tt hep-ph/9607289} (DIS96 proceedings).}
\nref\BV{J.~Bl\"umlein and A.~Vogt, in these proceedings.}
\nref\DAS{R.D.~Ball and S.~Forte, \PL\vyp{B335}{1994}{77};\vyp{B336}{1994}{77}.}
\nref\AFP{R.D.~Ball and S.~Forte, \PL\vyp{B405}{1997}{317}; 
{\tt hep-ph/9706459} (DIS97 proceedings).}

\noindent
The calculation of the Mellin transform of the Lipatov kernel
\eqn\pom{
\omega_P(\gamma)=\bar{\alpha}_s(\chi_0(\gamma)+\alpha_s
\chi_1(\gamma)),}
has recently been completed at NLLx.\cites{\FL,\CC} It was immediately
apparent that the correction to  $\omega_P(\half)$ is rather
large. Moreover for any reasonable value of
$\alpha_s$, the LLx minimum on the real axis at $\gamma = \half$
becomes a maximum (see fig.~1): at NLLx the asymptotic behaviour of 
cross-sections in the Regge limit is then determined not by the LLx
saddle point on the real axis, with small NLLx corrections, but 
rather by a complex pair of NLLx saddle points. Further expansion in
powers of $\alpha_s$ is thus not very helpful in the Regge limit.

In an attempt to circumvent these difficulties, we
here turn to the `$k_t$-factorization' approach,\cite\CH\ in which the Lipatov kernel is
used to deduce the LLx and NLLx terms in Altarelli-Parisi anomalous 
dimensions to all orders in $\alpha_s$. These can then be summed 
up, so that at NLLx we may write 
\midinsert
\vbox{\hbox{\hskip2.5truecm
\hfil\epsfxsize=6.5truecm\epsfysize=3truecm\epsfbox{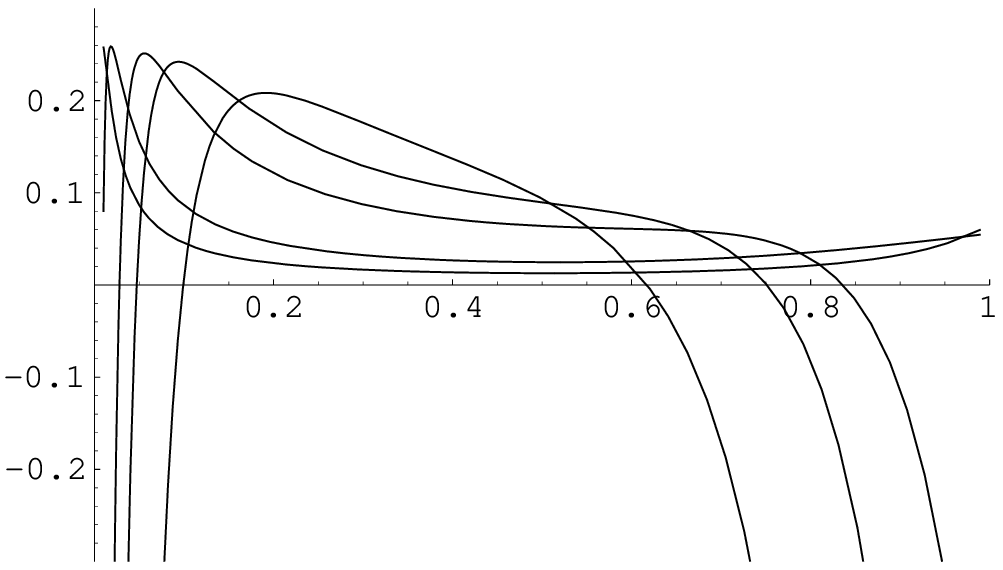}\hfil}
\vskip0.1truecm
\vskip-0.5truecm
\bigskip\noindent{\footnotefont\baselineskip6pt\narrower
Figure 1: The `pomeron intercept' eq.\pom\ 
at NLLx plotted against $\gamma$: the curves are for $\alpha_s
=0.1,0.05,0.03,0.01,0.005$ respectively (top to bottom in the middle).
}}
\medskip
\endinsert

\topinsert
\vskip-1.5truecm
\vbox{\hbox{\hskip-0.5truecm
\hfil\epsfxsize=6.5truecm\epsfbox{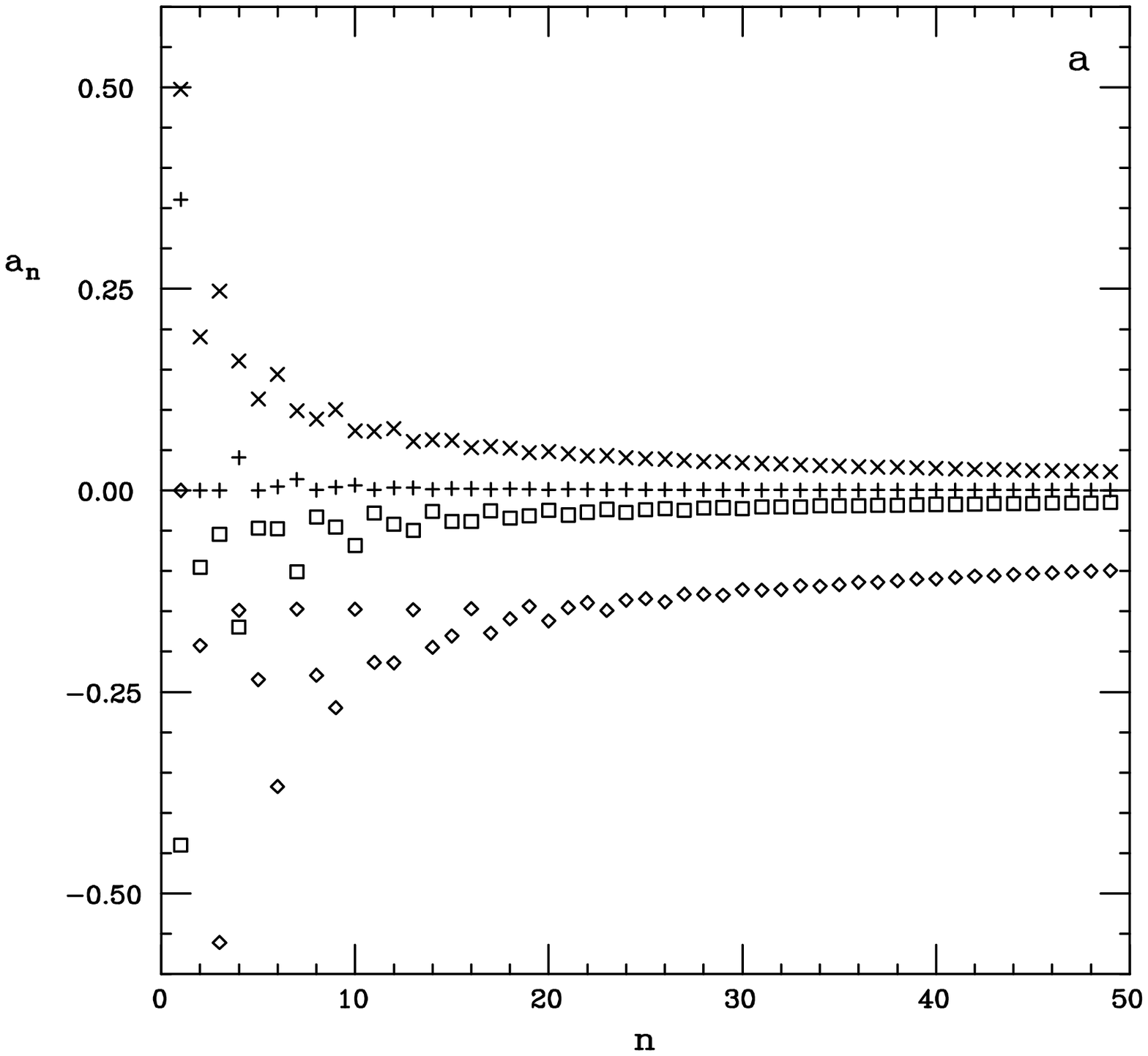}
\hskip-0.5truecm
\epsfxsize=6.5truecm\epsfbox{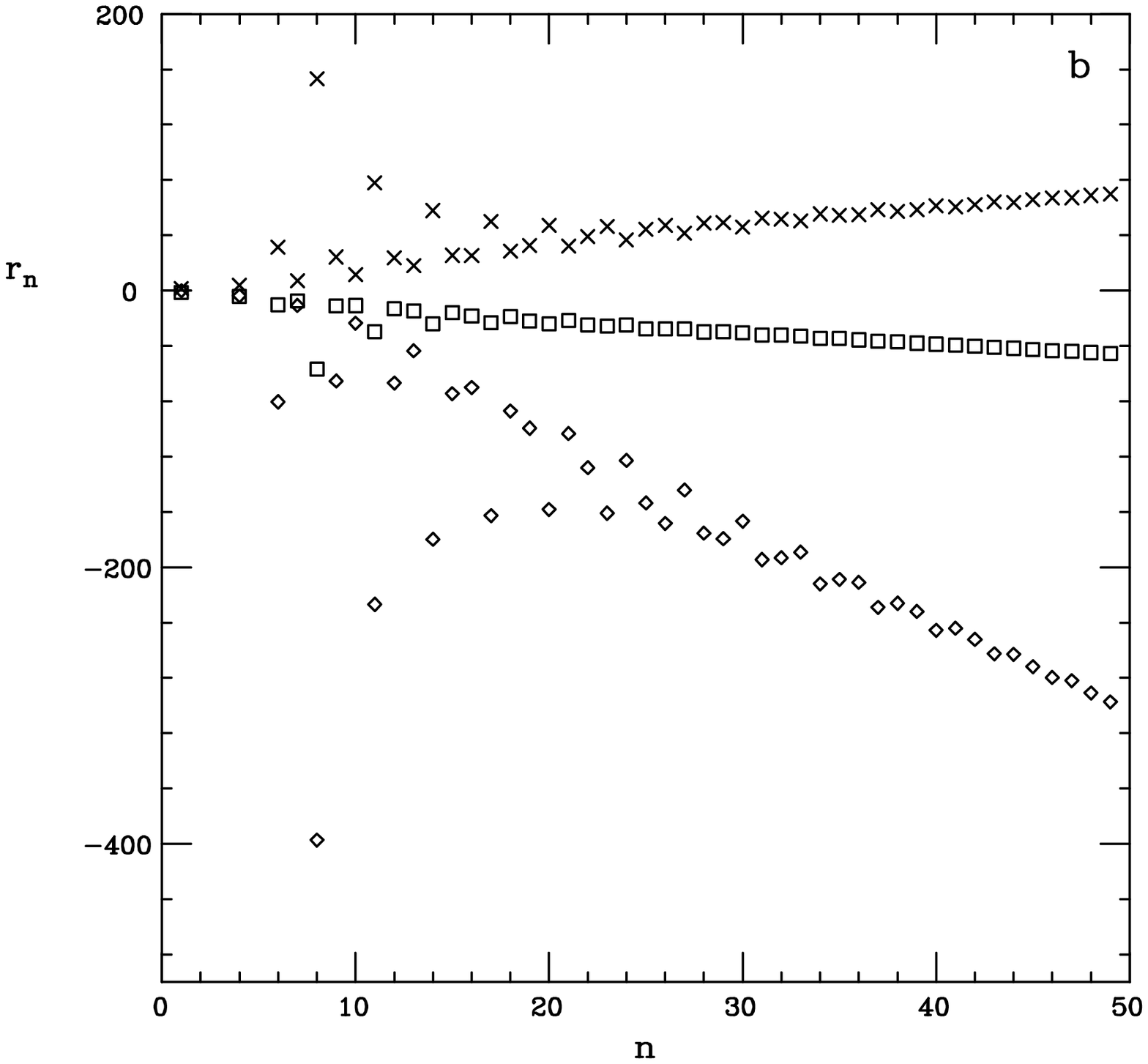}\hfil}
\vskip-2.0truecm
\bigskip\noindent{\footnotefont\baselineskip6pt\narrower
Figure 2: (a) coefficients $a_n\left(+\right)$,
$n_fb_n\left(\times\right)$, $c_n\left(\diamond\right)$, $n_fd_n\left(\square\right)$, 
 and (b) ratios 
$r_n= n_fb_n/a_n\left(\times\right)$, $c_n/a_n\left(\diamond\right)$, 
$n_fd_n/a_n\left(\square\right)$, all in the DIS scheme. We set
$n_f=6$ for definiteness. The ratios for $n=2,3,5$ are all infinite.}}
\medskip
\vskip-0.5truecm
\endinsert
\eqn\codef{
\gamma_{gg}=\hbox{$\sum_{n=1}^\infty$} 
(a_n+\alpha_s(c_n+n_f d_n))z^n,\qquad
\gamma_{qg}(N,\alpha_s)
=\alpha_s n_f \hbox{$\sum_{n=1}^\infty$}  b_n z^n,}
where $z\equiv 12\ln 2 \alpha_s /\pi N$, and the series converge for 
$|z|<1$.\cite\Summing

\topinsert
\vskip-1.5truecm
\vbox{\hbox{\hskip-0.5truecm
\hfil\epsfxsize=6.5truecm\epsfbox{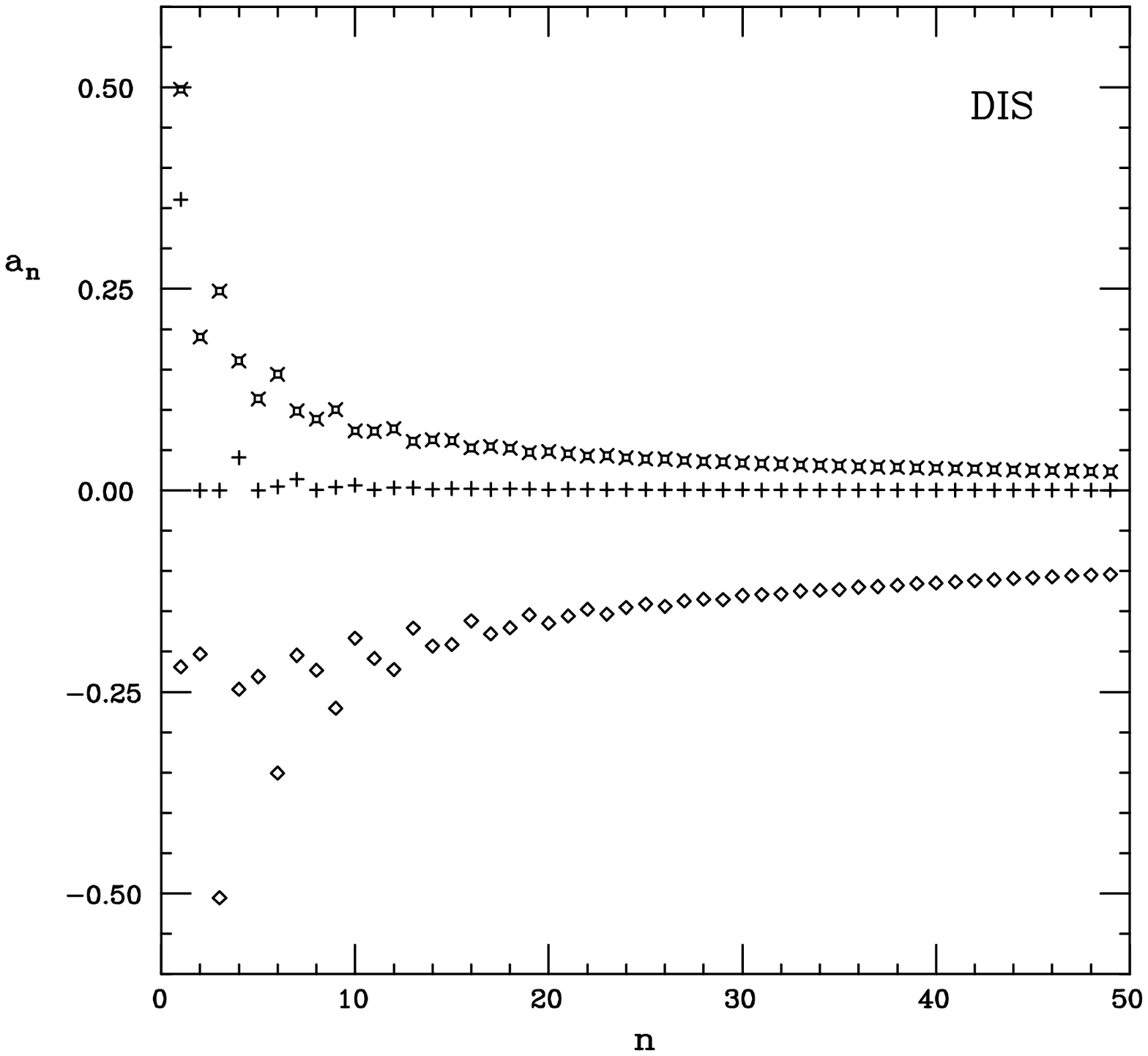}
\hskip-0.5truecm
\epsfxsize=6.5truecm\epsfbox{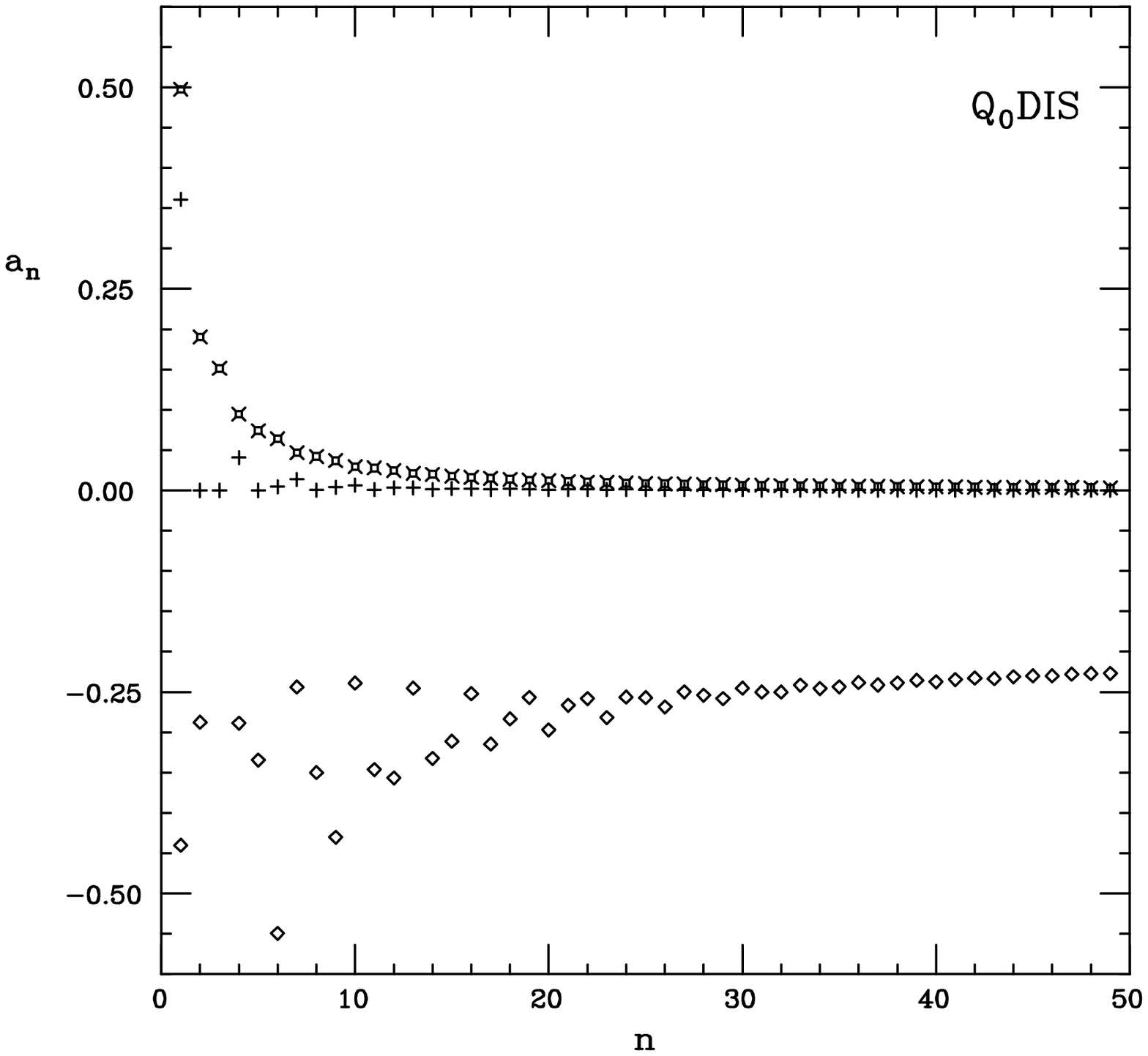}\hfil}
\vskip-3.0truecm
\hbox{\hskip-0.5truecm
\hfil\epsfxsize=6.5truecm\epsfbox{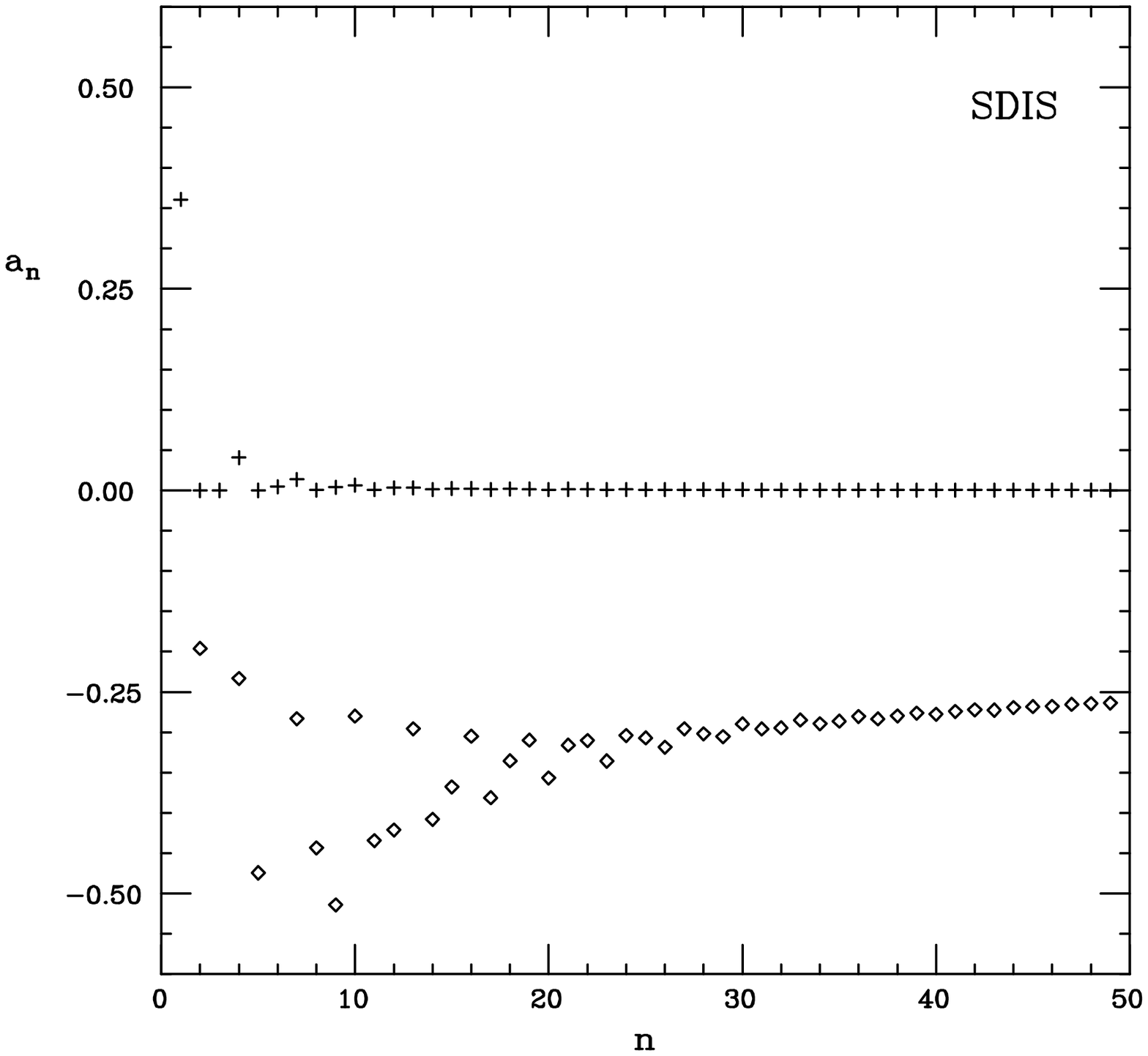}
\hskip-0.5truecm
\epsfxsize=6.5truecm\epsfbox{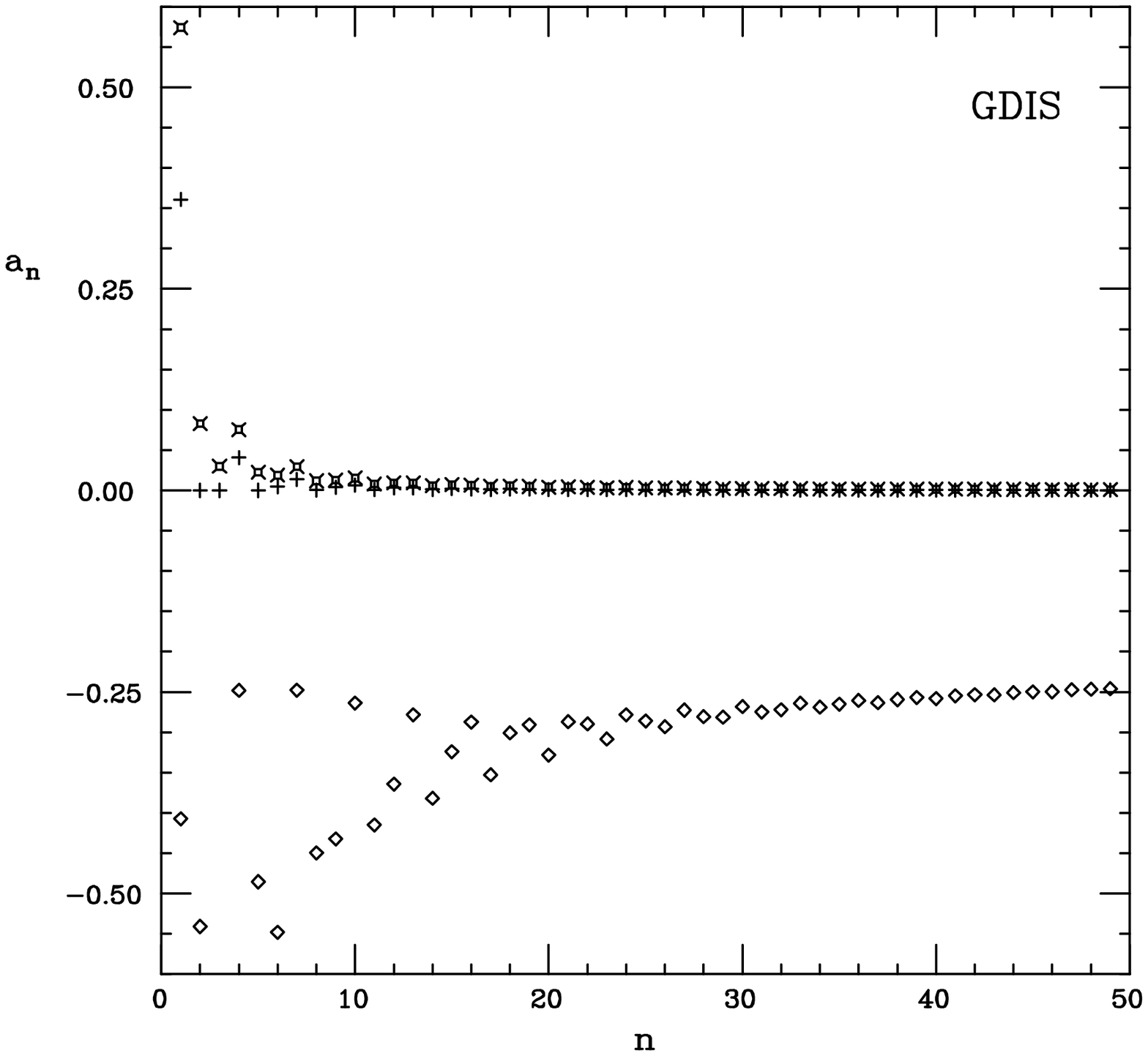}\hfil}
\vskip-3.0truecm
\hbox{\hskip-0.5truecm
\hfil\epsfxsize=6.5truecm\epsfbox{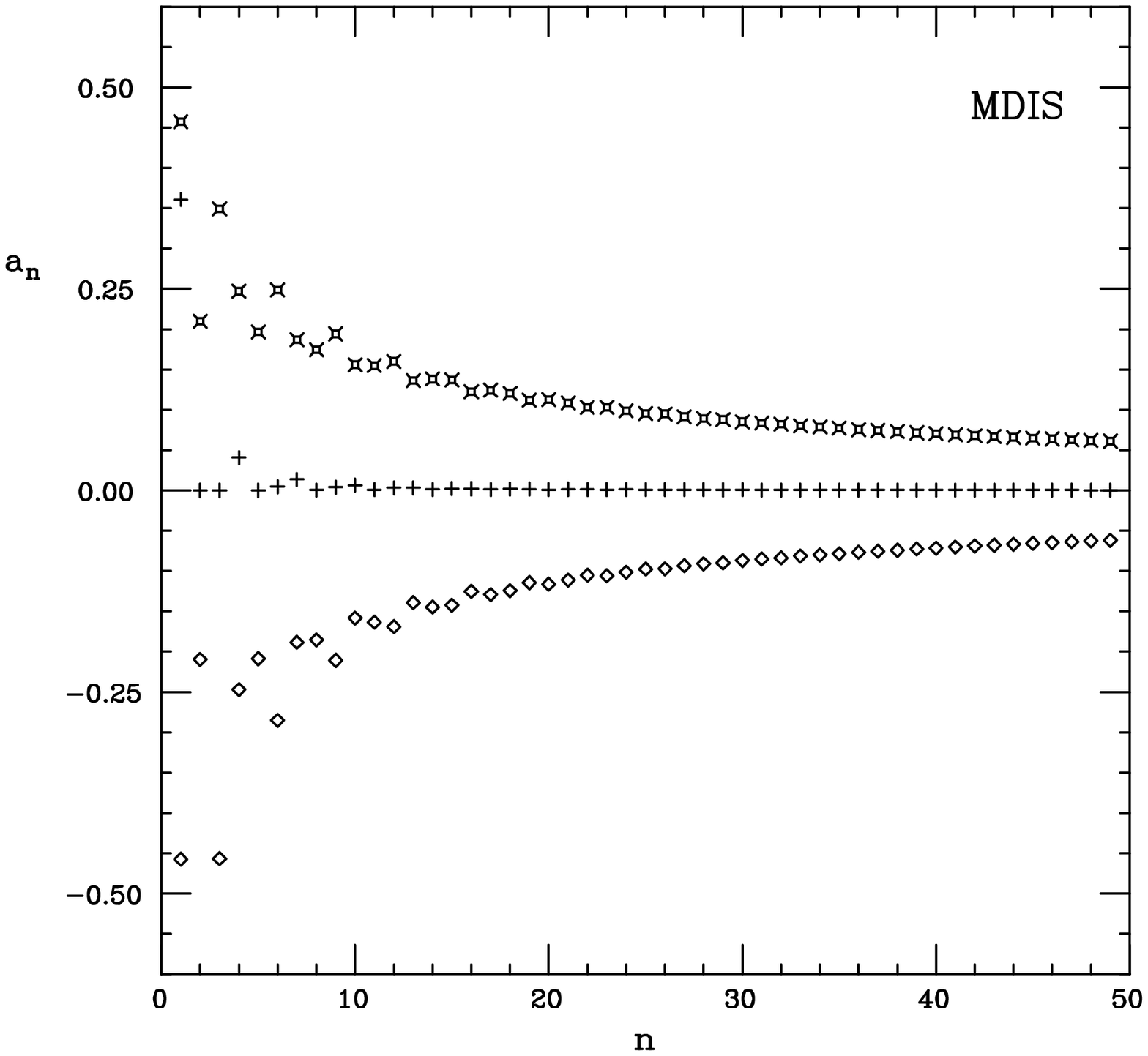}
\hskip-0.5truecm
\epsfxsize=6.5truecm\epsfbox{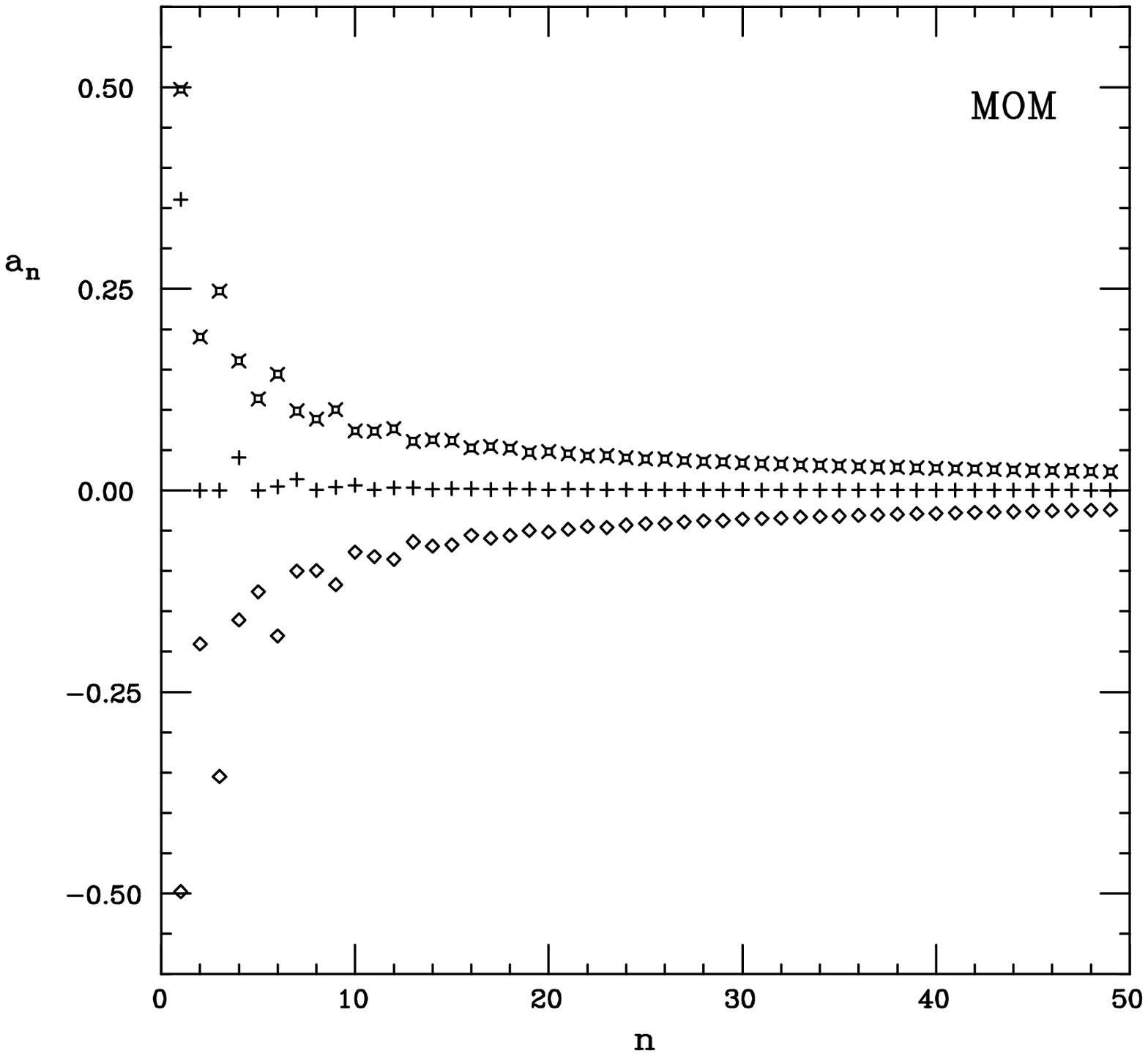}\hfil}
\vskip-2.0truecm
\bigskip\noindent{\footnotefont\baselineskip6pt\narrower
Figure 3: The coefficients  $a_n\left(+\right)$,
$n_f b_n\left(\times\right)$, $c_n+n_f d_n\left(\diamond\right)$,
 in six different factorization schemes.\cites{\FL,\CQz -\Mom} }}
\medskip
\endinsert

The NLLx coefficients $c_n$ and $d_n$ may be computed for all $n$ from
the result for $\chi_1(\gamma)$ in ref.1: the results are presented in
fig.~2. Although the purely gluonic NLLx coefficients $c_n$ are of order unity,
the LLx coefficients $a_n$, and the NLLx quark contributions $b_n$ and
$d_n$ are all very small. When the ratio of NLLx to LLx coefficients
is of order $\alpha_s^{-1}\sim 10$, perturbation theory is beginning
to break down. Here (fig.2b) the gluonic perturbative `corrections'
are simply enormous (and negative): $|d_n/a_n|\gg 10$ for all $n$. 
The failure of the perturbative expansion \codef\ is even more
dramatic than that of \pom. 

To explore whether this failure is due to an unfortunate choice of definition
of the gluon distribution, we consider various alternative LLx
factorization schemes (fig.~3). In schemes where $\gamma_{qg}$
is reduced (the $Q_0$-scheme,\cite\CQz\ the  SDIS
scheme\cite\SDIS\ in which $\gamma_{qg}=0$ by construction, and 
the `physical' scheme\cite\GDIS\ (GDIS)) the NLLx contribution to 
$\gamma_{gg}$ becomes even larger. In the MDIS scheme,\cite\Mom\ 
chosen to guarantee momentum conservation at NLLx, trouble is shared
between the quark and gluon sector, but the corrections are still very
large. Ratios of coefficients in all these schemes (and their $Q_0$
variants) are displayed in fig.~4. It is amusing that in all the
schemes the perturbative expansion gets systematically worse as $n$
increases (i.e. as $x$ decreases).

\topinsert
\vskip-1.5truecm
\vbox{\hbox{\hskip-0.5truecm
\hfil\epsfxsize=6.5truecm\epsfbox{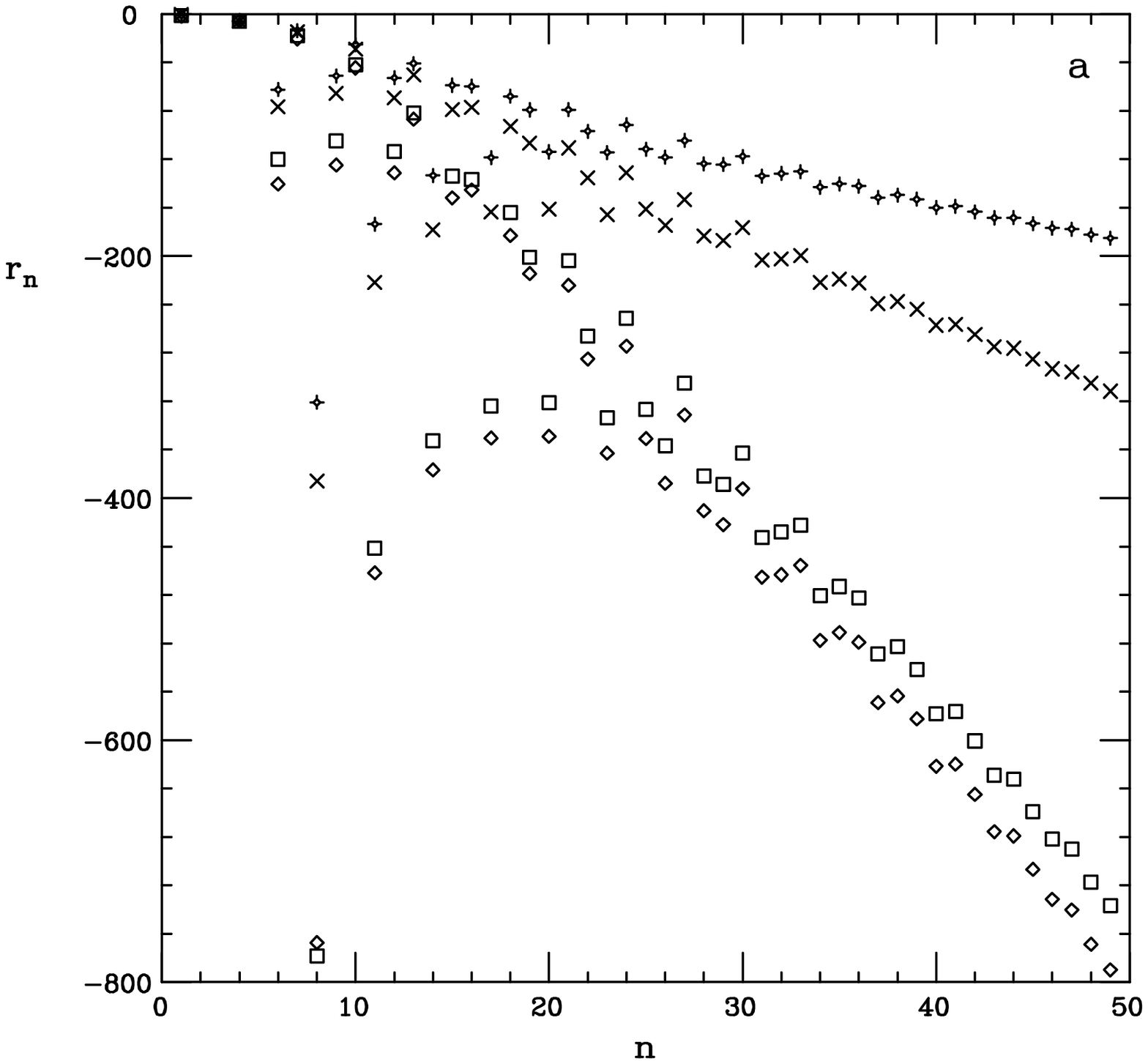}
\hskip-0.5truecm
\epsfxsize=6.5truecm\epsfbox{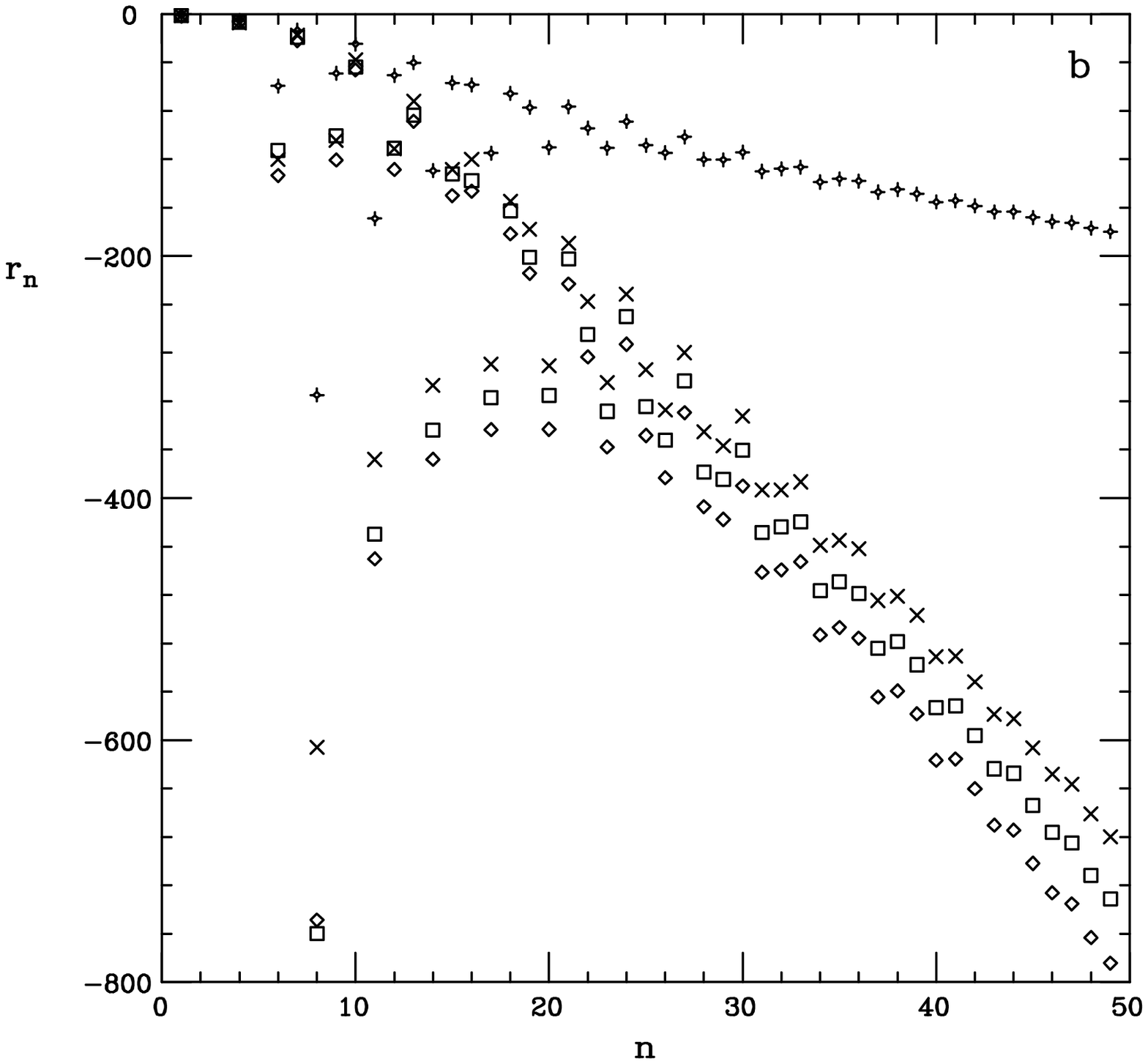}\hfil}
\vskip-2.0truecm
\bigskip\noindent{\footnotefont\baselineskip6pt\narrower
Figure 4: The ratios $(c_n+n_f d_n)/a_n$ in various factorization
schemes: a) DIS $\left(\times\right)$, SDIS\cite\SDIS\ $\left(\diamond\right)$
GDIS\cite\GDIS\ $\left(\square\right)$, MDIS\cite\Mom\ $\left(+\right)$ and 
b) in the corresponding $Q_0$ schemes.\cite\CQz}}
\medskip
\vskip-0.5truecm
\endinsert

Also in fig.~3 (MOM) we show the coefficients used in earlier fits to 1994
HERA data\cite\Rome\ which found that the data dislike the 
small-$x$ logarithms. Since these coefficients are all much smaller than 
those in the MDIS scheme, it follows immediately that HERA data
are completely incompatible with the NLLx logarithms. This is because 
such logs rapidly suppress the gluon distribution at small $x$ (the
NLLx corrections to $\gamma_{gg}$ are large and negative),
even driving the structure function negative in a linearised calculation.\cite\BV\

In conclusion, if the new NLLx calculations\cites{\FL,\CC} are correct, 
it is useless to sum logs of $x$ when evolving parton distributions or
structure functions: such calculations are of no phenomenological relevance since they 
rely on the false assumption that higher order corrections are small.
This marks a significant advance in our understanding: we now know
precisely why inclusive data show no sign of BFKL in the double
scaling\cite\DAS\ region. We believe that the development of a useful 
formulation of perturbative QCD in the
Regge limit requires a new approach to high energy factorization.\cite\AFP

\immediate\closeout\rfile\writestoppt
\bigskip
\noindent{{\bf References}}\smallskip{\frenchspacing%
\parindent=20pt
\ninepoint\baselineskip=11pt
\escapechar=` \input refs.tmp\vfill\eject}\nonfrenchspacing

\end